\documentstyle[bezier,12pt]{article}

\input epsf.sty

\newcommand{\insertfig}[2]{\mbox{\epsfxsize=#1cm \epsfbox{#2.eps}}}
\newcommand{\Asym}{\mathop{\mbox{\Large\bf A}}}
\newcommand{\Sym}{\mathop{\mbox{\Large\bf S}}}
\newcommand{\tAsym}{\mathop{\mbox{\bf A}}}
\newcommand{\tSym}{\mathop{\mbox{\bf S}}}
\newcommand{\g}{{\sl g}}
\newcommand{\ft}[2]{{\textstyle\frac{#1}{#2}}}

\begin{document}

\begin{titlepage}

\centerline{\large \bf Twist-three effects in two-photon processes.}

\vspace{15mm}

\centerline{\bf A.V. Belitsky$^{a,b}$, D. M\"uller$^b$}
\vspace{5mm}

\centerline{\it $^a$C.N.\ Yang Institute for Theoretical Physics}
\centerline{\it State University of New York at Stony Brook}
\centerline{\it NY 11794-3840, Stony Brook, USA}

\vspace{5mm}

\centerline{\it $^b$Institut f\"ur Theoretische Physik,
                Universit\"at Regensburg}
\centerline{\it D-93040 Regensburg, Germany}

\vspace{15mm}

\centerline{\bf Abstract}
\hspace{0.5cm}

We give a general treatment of twist-three effects in two-photon reactions.
We address the issue of the gauge invariance of the Compton amplitude in
generalized Bjorken kinematics and relations of twist-three `transverse'
skewed parton distributions to twist-two ones and interaction dependent
three-particle correlation functions. Finally, we discuss leading order
evolution of twist-three functions and their impact on the deeply virtual
Compton scattering.

\vspace{7cm}


\vspace{0.5cm}


\end{titlepage}

%%%%%%%%%%%%%%%%%%%%%%%%%%%%%%%%%%%%%%%%%%%%%%%%%%%%%%%%%%%%%%%%%%%%%
\section{Introduction.}
%%%%%%%%%%%%%%%%%%%%%%%%%%%%%%%%%%%%%%%%%%%%%%%%%%%%%%%%%%%%%%%%%%%%%

Exclusive two-photon reactions involving two hadrons (scattering or
production), like deeply virtual Compton scattering (DVCS)
\cite{MueRobGeyDitHor94,Ji97,Rad96}, $\gamma^\ast \gamma \to h h$ with small
invariant mass of the hadron system \cite{MueRobGeyDitHor94,DieGouPirTer98},
etc., are of special interest since they involve new nonperturbative
characteristics which generalize conventional parton densities and/or
distribution amplitudes. The leading twist factorization gives the
amplitude of these processes in terms of a perturbatively calculable
coefficient function and a skewed parton distribution (for DVCS)
\cite{Rad96,ColFre98,JiOsb98} or a generalized distribution amplitude (for
$\gamma^\ast\gamma \to hh$) \cite{Fre00} which are responsible for soft
physics. However, the twist-two analysis of the Compton scattering at
non-zero $t$-channel momentum transfer, $\Delta \neq 0$, is inadequate
since it violates explicitly the gauge invariance of the amplitude due to
approximation involved, $\Delta_\perp / \sqrt{- q^2} \ll 1$. This calls for
a consistent treatment of the effects suppressed in $\Delta_\perp$, i.e.\
twist expansion\footnote{Here we imply kinematical definition of twist.}.
As a first step one takes the contributions linear in $\Delta_\perp$ which
are of twist-three. Obviously, then the amplitude will be gauge invariant
to the twist-four accuracy. Repeating the steps one improves the amplitude
accordingly.

Let us demonstrate the peculiarities of the off-forward kinematics on
a simple example of a free Dirac fermion theory. As we will see, even
and odd parity structures `talk' to each other in the case at hand, when
the operators with total derivatives within the context of the operators
product expansion are relevant. This will be the source of restoration
of the electromagnetic gauge invariance of the two-photon amplitude
defined by a chronological product $T \left\{ j_\mu (x) j_\nu (y) \right\}$
of currents $j_\mu (x) = \bar\psi \gamma_\mu \psi$. The leading light-cone
singularity $(x - y)^2 \to 0$ arises from the hand-bag diagram and reads
\begin{eqnarray}
\label{OPE-naive}
T \left\{ j_\mu (x) j_\nu (y) \right\}
= i \bar\psi (x) \gamma_\mu /\!\!\!S (x - y) \gamma_\nu \psi (y)
+ i \bar\psi (y) \gamma_\nu /\!\!\!S (y - x) \gamma_\mu \psi (x) ,
\end{eqnarray}
where $/\!\!\!S (x) = \frac{1}{2 \pi^2} \frac{/\!\!\!x}{x^4}$ is
the free quark propagator. Taking into account the equation of motion
$/\!\!\!\partial \psi = 0$ and $/\!\!\!\partial \, /\!\!\!S (x)
= - i \delta (x)$ it is a simple task to show that the hand-bag diagram
respect current conservation. After performing the decomposition of
the Dirac structure in Eq.\ (\ref{OPE-naive}) it reduces to\footnote{We
use the conventions for Dirac and Lorentz tensors from Itzykson and
Zuber \cite{ItzZub80}.}
\begin{eqnarray}
\label{OPE-naive-decom}
T \left\{ j_\mu (x) j_\nu (y) \right\}
\!\!\!&=&\!\!\! S_{\mu \nu; \rho \sigma} i S_\rho (x - y)
\left\{
\bar\psi (x) \gamma_\sigma \psi (y) - ( x \leftrightarrow y )
\right\} \\
&-&\!\!\! i \epsilon_{\mu\nu\rho\sigma} i S_\rho (x - y)
\left\{
\bar{\psi}(x) \gamma_\sigma \gamma_5 \psi (y) + ( x \leftrightarrow y )
\right\} , \nonumber
\end{eqnarray}
where $S_{\mu \nu; \rho \sigma} = g_{\mu\rho} g_{\nu\sigma} + g_{\mu\sigma}
g_{\nu\rho} - g_{\mu\nu} g_{\rho\sigma}$. One finds that current
conservation does not separately occur in both terms. Employing the
density matrix $|P_1 \rangle \langle P_2 |$ for free fields, we find,
of course, that the amplitude $ \langle P_2 | T \left\{ j_\mu (x) j_\nu
(y) \right\} |P_1 \rangle$ respects current conservation, however, there
appears a cancellation between both matrix elements of different parity on 
the r.h.s.\ in Eq.\ (\ref{OPE-naive-decom}). So we realize that the current
conservation is not manifest in the decomposition (\ref{OPE-naive-decom}).
This arises due to operators with total derivatives in twist
decomposition of $\bar\psi (x) \gamma_\mu (1, \gamma_5) \psi (y)$, see
later Eq.\ (\ref{R3-part}). However, once this decomposition is performed 
the gauge invariance is restored automatically. As compared to this simple 
example the only complication in QCD to leading order emerges due to the
presence of the interaction dependent three-particle contributions. This 
does not present any difficulty and will be solved in the next section.

Our consequent presentation is organized as follows. The next section is
devoted to a detailed description of the twist-three formalism for the
generalized Compton amplitude and as a result the restoration of the gauge
invariance sketched above. In section \ref{TwistDecomp} we address the issue
of a twist decomposition of light-ray (and local) operators in a situation
when total derivatives are relevant. Then we turn in section 4 to the two
cases of matrix elements of operators sandwiched between spin-0 and $\ft12$
hadrons and derive relations for twist-three two-particle `transverse'
functions in terms of twist-two (Wandzura-Wilczek contribution) and
three-particle correlation functions. In section 5 we comment on the
phenomenological consequences for different asymmetries measurable in the
DVCS process, and then in section \ref{Evolution} point out that the 
evolution of twist-three skewed distributions is known in leading 
logarithmic approximation from analogous studies for forward kinematics. 
Finally, we give our conclusions.

%%%%%%%%%%%%%%%%%%%%%%%%%%%%%%%%%%%%%%%%%%%%%%%%%%%%%%%%%%%%%%%%%%%%%
\section{Generalized two-photon amplitude.}
\label{Amplitude}
%%%%%%%%%%%%%%%%%%%%%%%%%%%%%%%%%%%%%%%%%%%%%%%%%%%%%%%%%%%%%%%%%%%%%

The amplitude of scattering of two virtual photons with momenta $q_1$
and $q_2$ on a hadron target, with incoming (outgoing) momentum $P_1$
($P_2$), is given by a Fourier transform of a correlator of
electromagnetic currents,
\begin{equation}
\label{ComptonAmplit}
T_{\mu\nu} = i \int d^4 x\, e^{i q \cdot x}
\langle P_2 | T \left\{ j_\mu (x/2) j_\nu (- x/2) \right\} | P_1 \rangle .
\end{equation}
One introduces the vectors $q = \ft12 (q_1 + q_2)$ and $P = P_1 + P_2$,
$\Delta = P_2 - P_1 = q_1 - q_2$ to describe the amplitude. They can be
used to construct a pair of the light-cone vectors $n_\mu$ and $n^\star_\mu$,
such that $n^2 = n^{\star 2} = 0$ and $n \cdot n^\star = 1$, as follows
\begin{equation}
n_\mu = - \frac{2 \xi}{q^2 \sqrt{1 - 4 (\xi \delta)^2}} \, q_\mu
- \frac{1 - \sqrt{1 - 4 (\xi \delta)^2}}{2 q^2 \delta^2
\sqrt{1 - 4 (\xi \delta)^2}} \, P_\mu , \quad\
n^\star_\mu = \frac{\xi \delta^2}{\sqrt{1 - 4 (\xi \delta)^2}} \, q_\mu
+ \frac{1 + \sqrt{1 - 4 (\xi \delta)^2}}{
4 \sqrt{1 - 4 (\xi \delta)^2}} \, P_\mu ,
\end{equation}
where $\delta^2 \equiv (M^2 - \Delta^2/4)/q^2$. The transverse metric and
antisymmetric tensor are defined as, $g_{\mu\nu}^\perp \equiv g_{\mu\nu}
- n_\mu n^\star_\nu - n_\nu n^\star_\mu$, $\epsilon^\perp_{\mu\nu}
\equiv \epsilon_{\mu \nu - +}$. Here and in the following we use the
conventions for the generalized Bjorken variable $\xi = - q^2/ P \cdot q$
and skewedness $\eta = \Delta \cdot q / P \cdot q$ which are scaling
variables of (\ref{ComptonAmplit}). Neglecting the corrections
${\cal O}(\delta^2)$ we have then to twist-four accuracy
\begin{equation}
n_\mu = - \frac{\xi}{q^2} (2 q_\mu + \xi P_\mu), \qquad
n^\star_\mu = \ft12 P_\mu , \qquad
\Delta_\mu = 2 \eta \, n^\star_\mu + \Delta^\perp_\mu .
\end{equation}
This approximation will be used throughout in our analysis.

%%%%%%%%%%%%%%%%%%%%%%%%%%%%%%%%%%%%%%%%%%%%%%%%%%%%%%%%%%%%%%%%%%%%%
%                           Figure 1                                %
%%%%%%%%%%%%%%%%%%%%%%%%%%%%%%%%%%%%%%%%%%%%%%%%%%%%%%%%%%%%%%%%%%%%%
\begin{figure}[t]
\begin{center}
\hspace{0cm}
\mbox{
\begin{picture}(0,50)(100,0)
\put(-50,-14){\insertfig{10}{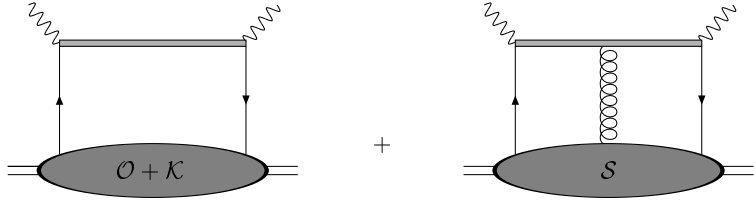}}
\end{picture}
}
\end{center}
\caption{\label{handbagdiagram} Two- and three-leg coefficient functions
(plus crossed diagrams) for DVCS process which form a gauge invariant
amplitude to twist-three accuracy.}
\end{figure}
%%%%%%%%%%%%%%%%%%%%%%%%%%%%%%%%%%%%%%%%%%%%%%%%%%%%%%%%%%%%%%%%%%%%%

In the generalized Bjorken kinematics $-q^2 \to \infty$, $P \cdot q \to
\infty$, $\Delta \cdot q \to \infty$, $\Delta^2 = {\rm finite}$, $\xi$ and
$\eta$ being fixed the contribution to (\ref{ComptonAmplit}) comes from the
diagrams in Fig.\ \ref{handbagdiagram}
\begin{equation}
T_{\mu\nu} = T_{\mu\nu}^2 + T_{\mu\nu}^3 ,
\end{equation}
where $T_{\mu\nu}^2$ comes from hand-bag diagram (left) while
$T_{\mu\nu}^3$ from antiquark-gluon-quark one (right). Explicit
calculation, done in the light cone gauge $B_+ = 0$, gives
\begin{eqnarray}
\label{2Part}
T_{\mu\nu}^2 = - \frac{1}{2 q^2} \int dx
\Bigg\{ \!\!\!\!\!&&\!\!\!\!\!
C^{(+)} (x, \xi)
\left[
S_{\mu \nu; \rho -} \, {^V\!K_\rho} (x, \eta)
+ 2 i \epsilon_{\mu \nu \rho \sigma} q_\sigma {^A O_\rho} (x, \eta)
\right] \nonumber\\
-\!\!\!\!&&\!\!\!\!\!
C^{(-)} (x, \xi)
\left[
\frac{q^2}{\xi} S_{\mu \nu; \rho +} \, {^V\!\!O_\rho} (x, \eta)
- i \epsilon_{\mu \nu \rho -}
\left( 2 x \, {^A O_\rho} (x, \eta) - {^A K_\rho} (x, \eta) \right)
\right]
\Bigg\} ,
\end{eqnarray}
for the first one, and
\begin{eqnarray}
T_{\mu\nu}^3 = - \frac{1}{2 q^2} \int dx \int \frac{d\tau}{\tau - i 0}
\Bigg\{ \!\!\!\!\!&&\!\!\!\!\!
C^{(+)} (x, \xi) S_{\mu \nu; \rho -}
\left[ {^V\!\!S^{-}_\rho} (x, x - \tau, \eta)
+ {^V\!\!S^{+}_\rho} (x + \tau, x, \eta) \right] \nonumber\\
+\!\!\!\!&&\!\!\!\!\!
C^{(-)} (x, \xi) S_{\mu \rho; \nu -}
\left[ {^V\!\!S^{-}_\rho} (x, x - \tau, \eta)
- {^V\!\!S^{+}_\rho} (x + \tau, x, \eta) \right]
\Bigg\} ,
\end{eqnarray}
for the second. Here $S_{\mu \nu; \rho \sigma}$ tensor defined in the
introduction and tree coefficient functions are
\begin{equation}
C^{(\pm)} (x, \xi)
= (1 - x/\xi - i 0)^{- 1} \pm (1 + x/\xi - i 0)^{- 1} .
\end{equation}
We use here the following conventions for generalized functions
\begin{eqnarray}
&&{^I\!O_\rho} (x, \eta)
= \int \frac{d\kappa}{2 \pi} {\rm e}^{i \kappa x} \
\langle P_2 |
{^I\! {\cal O}_\rho} \left( \ft{\kappa}2 , - \ft{\kappa}2 \right)
| P_1 \rangle ,
\quad
{^I\!K_\rho} (x, \eta)
= \int \frac{d\kappa}{2 \pi} {\rm e}^{i \kappa x} \
\langle P_2 |
{^I\!{\cal K}_\rho} \left( \ft{\kappa}2, - \ft{\kappa}2 \right)
| P_1 \rangle , \nonumber\\
&&{^I\! S^{\pm}_\rho} (x_1, x_2, \eta)
= \int \frac{d\kappa_1}{2 \pi} \frac{d\kappa_2}{2 \pi}
{\rm e}^{\ft{i}{2} \kappa_1 (x_1 + x_2) + i \kappa_2 (x_1 - x_2)} \
\langle P_2 |
{^I\!{\cal S}^\pm_\rho}
\left( \ft{\kappa_1}2, \kappa_2, - \ft{\kappa_1}2 \right)
| P_1 \rangle ,
\label{MFSdef}
\end{eqnarray}
where the two-quark operators are defined by
\begin{equation}
\label{Operator}
{^I\! {\cal O}_\rho} (\kappa, - \kappa)
= \bar\psi (- \kappa n) {^I\!{\mit\Gamma}_\rho} \psi (\kappa n) ,
\qquad
{^I\!{\cal K}_\rho} (\kappa, - \kappa)
= \bar\psi (- \kappa n)\, {^I\!{\mit\Gamma}_+} i\!
\stackrel{\leftrightarrow}{\partial}\!\!{}^\perp_\rho \,
\psi (\kappa n) ,
\end{equation}
with ${^I\!{\mit\Gamma}} = \{\gamma_\rho, \gamma_\rho \gamma_5$\}
for  $I = V, A$ and antiquark-gluon-quark ones read
\begin{eqnarray}
{^V\!\!{\cal S}^{\pm}_\rho} (\kappa_1, \kappa_2, \kappa_3)
&=& i \g \bar\psi (\kappa_3 n)
\left[
\gamma_+ G_{+ \rho} (\kappa_2 n)
\pm
i \gamma_+ \gamma_5 {\widetilde G}_{+ \rho} (\kappa_2 n)
\right]
\psi (\kappa_1 n) , \nonumber\\
{^A\!{\cal S}^{\pm}_\rho} (\kappa_1, \kappa_2, \kappa_3)
&=& i \g \bar\psi (\kappa_3 n)
\left[
\gamma_+ \gamma_5 G_{+ \rho} (\kappa_2 n)
\pm
i \gamma_+ {\widetilde G}_{+ \rho} (\kappa_2 n)
\right]
\psi (\kappa_1 n) .
\end{eqnarray}
The latter two are related to each other by the `duality' equation
\begin{equation}
\label{Pro-Dual}
i \epsilon^\perp_{\rho\sigma} \, {^A\!{\cal S}^{\pm}_\sigma}
= \pm \, {^V\!\!{\cal S}^{\pm}_\rho} .
\end{equation}

By means of the quark equation of motion, $\not\!\!\!{\cal D} \psi = 0$,
we can obtain the following relation between the correlation functions
introduced so far
\begin{eqnarray}
\frac{\partial}{\partial \kappa}
{^V\!\!{\cal O}^\perp_\rho} \left( \kappa , - \kappa \right)
\!\!\!&-&\!\!\! i \epsilon^\perp_{\rho \sigma} \partial_+
{^A {\cal O}^\perp_\sigma} \left( \kappa , - \kappa \right)
+ i {^V\! {\cal K}_\rho} \left( \kappa , - \kappa \right)
+ i \epsilon^\perp_{\rho \sigma} \partial^\perp_\sigma
{^A {\cal O}_+} \left( \kappa , - \kappa \right) \nonumber\\
&+&\!\!\! \int d \lambda
\left\{
w (\lambda - \kappa) {^V\!{\cal S}^-_\rho}
\left( \kappa , \lambda, - \kappa \right)
+
w (\lambda + \kappa) {^V\!{\cal S}^+_\rho}
\left( \kappa , \lambda, - \kappa \right)
\right\} = 0, \nonumber\\
\partial_+
{^V\!\!{\cal O}^\perp_\rho} \left( \kappa , - \kappa \right)
\!\!\!&-&\!\!\! i \epsilon^\perp_{\rho \sigma}
\frac{\partial}{\partial \kappa}
{^A {\cal O}^\perp_\sigma} \left( \kappa , - \kappa \right)
+ \epsilon^\perp_{\rho \sigma} {^A\! {\cal K}_\sigma}
\left( \kappa , - \kappa \right)
- \partial^\perp_\rho {^V\!\!{\cal O}_+}
\left( \kappa , - \kappa \right) \nonumber\\
&+&\!\!\! \int d \lambda
\left\{
w (\lambda - \kappa) {^V\!{\cal S}^-_\rho}
\left( \kappa , \lambda, - \kappa \right)
-
w (\lambda + \kappa) {^V\!{\cal S}^+_\rho}
\left( \kappa , \lambda, - \kappa \right)
\right\} = 0 ,
\end{eqnarray}
where $w (\kappa) = - \theta (\kappa)$ for the ML prescription on the
infrared pole in the gluon propagator. Here $\partial$ stands for the
total derivative. In terms of the momentum fraction space functions
introduced in Eqs.\ (\ref{MFSdef}) these relations read
\begin{eqnarray}
2 x \, {^V\!\!O^\perp_\rho} (x, \eta)
+
2 \eta \, i \epsilon^\perp_{\rho \sigma} {^A O^\perp_\sigma} (x, \eta)
\!\!\!&-&\!\!\! i
\epsilon^\perp_{\rho \sigma} \Delta^\perp_\sigma \, {^A O_+} (x, \eta)
- {^V\!\!K^\perp_\rho} (x, \eta) \nonumber\\
&-&\!\!\! \int \frac{d \tau}{\tau - i 0}
\left\{
{^V\!\! S^-_\rho} (x, x - \tau, \eta)
+
{^V\!\! S^+_\rho} (x + \tau, x, \eta)
\right\} = 0, \nonumber\\
2 x \, i \epsilon^\perp_{\rho \sigma} {^A O^\perp_\sigma} (x, \eta)
+
2 \eta \, {^V\!\!O^\perp_\rho} (x, \eta)
\!\!\!&-&\!\!\! \Delta^\perp_\rho \, {^V\!\!O_+} (x, \eta)
- i \epsilon^\perp_{\rho \sigma} {^A\!K^\perp_\sigma} (x, \eta) \nonumber\\
&+&\!\!\! \int \frac{d \tau}{\tau - i 0}
\left\{
{^V\!\! S^-_\rho} (x, x - \tau, \eta)
-
{^V\!\! S^+_\rho} (x + \tau, x, \eta)
\right\} = 0.
\end{eqnarray}
So that expressing $K$ in terms of other correlation functions in
Eq.\ (\ref{2Part}) we get the result
\begin{eqnarray}
\label{GaugeInvAmpl}
T_{\mu\nu} \!\!\!&=&\!\!\! - \frac{1}{q^2} \int dx \,
\Bigg\{
{\cal T}^{(1)}_{\mu\nu} \ C^{(-)} (x, \xi) \, {^V\!\!O_+} (x, \eta)
+
{\cal T}^{(2)}_{\mu\nu} \ C^{(+)} (x, \xi) \, {^A O_+} (x, \eta)
\nonumber\\
&&\qquad\qquad+
{\cal T}^{(3)}_{\mu\nu;\rho} \ C^{(-)} (x, \xi) \,
{^V\!\!O^\perp_\rho} (x, \eta)
+
{\cal T}^{(4)}_{\mu\nu;\rho} \ C^{(+)} (x, \xi) \,
{^A O^\perp_\rho} (x, \eta)
\Bigg\} ,
\end{eqnarray}
where the support properties of distributions restrict the integration
range of the variable $x$ within the interval $[-1, 1]$. Here the Lorentz
tensors are
\begin{eqnarray}
&&{\cal T}^{(1)}_{\mu\nu} =
n^\star_\mu \left( q_\nu + \xi n^\star_\nu - \ft12 \Delta^\perp_\nu \right)
+
n^\star_\nu \left( q_\mu + \xi n^\star_\mu + \ft12 \Delta^\perp_\mu \right)
- q \cdot n^\star g_{\mu\nu} ,
\nonumber\\
&&{\cal T}^{(2)}_{\mu\nu} =
i \epsilon_{\mu \nu - \sigma} q_\sigma
- \ft{i}2 \epsilon^\perp_{\rho \sigma} \Delta^\perp_\sigma
\left( n^\star_\mu g_{\rho\nu} + n^\star_\nu g_{\rho\mu} \right) ,
\nonumber\\
&&{\cal T}^{(3)}_{\mu\nu;\rho} =
\left( q_\nu + ( 2 \xi - \eta ) n^\star_{\nu\phantom{\mu}} \right)
g_{\mu \rho}
+
\left( q_\mu + ( 2 \xi + \eta ) n^\star_\mu \right) g_{\nu \rho} ,
\nonumber\\
&&{\cal T}^{(4)}_{\mu\nu;\rho} =
i \epsilon_{\mu \nu \rho \sigma} q_\sigma
- i \eta \epsilon^\perp_{\rho \sigma}
\left( n^\star_\mu \, g_{\sigma\nu} + n^\star_\nu \, g_{\sigma\mu} \right) .
\end{eqnarray}
These expressions are obviously target independent. In the case of scalar
target our result reduces to the one obtained in Ref.\ \cite{AniPirTer00}.

It is an easy exercise to check that, after accounting for twist-three
corrections, gauge invariance is fulfilled up to ${\cal O} (\Delta_\perp^2)$,
i.e.\ twist-four effects: $q_{1 \nu} {\cal T}^{(i)}_{\mu\nu} = q_{2 \mu}
{\cal T}^{(i)}_{\mu\nu} = {\cal O} (\Delta_\perp^2)$, where we have used
Sudakov decomposition for the photon momenta $q_{1 \mu} = - \frac{q^2}{2\xi}
n_\mu - (\xi - \eta) n^\star_\mu + \ft12 \Delta^\perp_\mu$ and $q_{2 \mu}
= - \frac{q^2}{2\xi} n_\mu - (\xi + \eta) n^\star_\mu - \ft12
\Delta^\perp_\mu$. Interesting to note that the last two structures
${\cal T}_{\mu \nu; \rho}^{(i)}$ are gauge invariant provided we
simultaneously contract them both with incoming and outgoing photons
$q_{1 \nu} q_{2 \mu} {\cal T}^{(i)}_{\mu\nu; \rho} = 0$. Namely, e.g.\
for the last tensor we get $q_{1 \nu} {\cal T}^{(4)}_{\mu \nu; \rho} =
- \ft12 \xi \, {\cal P}_{\mu\nu} P_\nu \widetilde\Delta^\perp_\rho$ and
$q_{2 \mu} {\cal T}^{(4)}_{\mu \nu; \rho} = - \ft12 \xi \, P_\mu
{\cal P}_{\mu\nu} \widetilde\Delta^\perp_\rho$, where we use
throughout the convention $\widetilde\Delta_\rho^\perp \equiv i
\epsilon^\perp_{\rho \sigma} \Delta_\rho$ and introduced a projector
${\cal P}_{\mu\nu} = g_{\mu\nu} - q_{1 \mu} q_{2 \nu}/ q_1 \cdot q_2$
which gives $\xi \, P_\mu {\cal P}_{\mu\nu} = q_\nu + \ft12 (2 \xi -
\eta) P_\nu$ and $\xi \, {\cal P}_{\mu\nu} P_\nu = q_\mu + \ft12 (2
\xi + \eta) P_\mu$ up to terms of order ${\cal O} (\Delta_\perp)$
which we drop since they would exceed the accuracy we work to. A similar
result we get for ${\cal T}^{(3)}_{\mu \nu; \rho}$ with
$\widetilde\Delta_\perp$ being replaced by $\Delta_\perp$. Using the
definitions of the photon and hadron momenta in terms of light-cone
vectors, consequent simple algebra leads to the following structures
in photon and hadron momenta
\begin{eqnarray}
\label{TensorStructure}
&&{\cal T}^{(1)}_{\mu \nu} = \frac{q^2}{2 \xi}
\left(
g_{\mu\nu} - \frac{q_{1 \mu} \, q_{2 \nu}}{q_1 \cdot q_2}
\right)
+ \frac{\xi}{2}
\left(
P_\mu - \frac{P \cdot q_2}{q_1 \cdot q_2} \, q_{1 \mu}
\right)
\left(
P_\nu - \frac{P \cdot q_1}{q_1 \cdot q_2} \, q_{2 \nu}
\right) , \nonumber\\
&&{\cal T}^{(2)}_{\mu\nu}
= \ft{i}2 \epsilon_{\theta \lambda \rho \sigma} P_\rho \, q_\sigma
\left(
g_{\mu \theta} - \frac{P_\mu \, q_{2 \theta}}{P \cdot q_2}
\right)
\left(
g_{\nu \lambda} - \frac{P_\nu \, q_{1 \lambda}}{P \cdot q_1}
\right) , \nonumber\\
&&{\cal T}^{(3)}_{\mu \nu; \rho}
= \xi \left(
g_{\mu \rho} - \frac{q_{1 \mu} \, q_{2 \rho}}{q_1 \cdot q_2}
\right)
\left(
P_\nu - \frac{P \cdot q_1}{q_1 \cdot q_2} \, q_{2 \nu}
\right)
+ \xi \left(
g_{\nu \rho} - \frac{q_{1 \rho} \, q_{2 \nu}}{q_1 \cdot q_2}
\right)
\left(
P_\mu - \frac{P \cdot q_2}{q_1 \cdot q_2} \, q_{1 \mu}
\right) , \nonumber\\
&&{\cal T}^{(4)}_{\mu \nu; \rho}
= i \epsilon_{\theta \lambda \rho \sigma} q_\sigma
\left(
g_{\mu \theta} - \frac{P_\mu \, q_{2 \theta}}{P \cdot q_2}
\right)
\left(
g_{\nu \lambda} - \frac{P_\nu \, q_{1 \lambda}}{P \cdot q_1}
\right) ,
\end{eqnarray}
where a missing twist-four part is restored minimally according to previous
discussion. All structures have a well defined forward limit. It will be
shown in section \ref{suppressionDVCS} that these terms are equivalent to
those used in our previous studies \cite{BelSch98,BelMulNieSch00}. In
subsequent sections we derive relations between the `transverse' generalized
distributions introduced here and conventional twist-two ones as well as
three-particle correlation functions.

%%%%%%%%%%%%%%%%%%%%%%%%%%%%%%%%%%%%%%%%%%%%%%%%%%%%%%%%%%%%%%%%%%%%%
\section{Twist decomposition.}
\label{TwistDecomp}
%%%%%%%%%%%%%%%%%%%%%%%%%%%%%%%%%%%%%%%%%%%%%%%%%%%%%%%%%%%%%%%%%%%%%

In this section we present a decomposition of the two-quark operators into
separate twist components to twist-three accuracy. Similar analyses have
been done in Refs.\ \cite{BalBra89,BalBraKoiTan98,GeyLaz99}. However, an
essential new ingredient of our study is the treatment of operators with
total derivatives. Since the group-theoretical notion of twist as the
dimension minus spin of an operator is well defined for local operators, our
strategy will be thus: first, an expansion of a light cone operator in
infinite Taylor series in local ones, second, an extraction of definite
twist components and as a final step the resummation of the result back into
nonlocal form. An alternative approach directly based in terms of light-ray
operators is described in Ref.\ \cite{BalBraKoiTan98}. In the following we
will not take care of trace terms, proportional to $n_\rho$, since they
only contribute at twist-four level.

The Taylor expansion of Eq.\ (\ref{Operator}) in terms of local operators
simply reads
\begin{equation}
\label{Def-Ope}
{\cal O}_\rho (\kappa, - \kappa)
= \sum_{j = 0}^{\infty} \frac{(-i \kappa)^j}{j!}
n_{\mu_1} \dots n_{\mu_j} {\cal O}_{\rho; \mu_1 \dots \mu_j} ,
\qquad\mbox{with}\qquad
{\cal O}_{\rho; \mu_1 \dots \mu_j} =
\bar\psi {\mit\Gamma}_\rho \,
i\!\stackrel{\leftrightarrow}{\cal D}_{\mu_1}
\dots
i\!\stackrel{\leftrightarrow}{\cal D}_{\mu_j}
\psi ,
\end{equation}
where $\stackrel{\leftrightarrow}{\cal D}_{\mu} \ = \
\stackrel{\rightarrow}{\cal D}_{\mu}
- \stackrel{\leftarrow}{\cal D}_{\mu}$ and ${\cal D}_\mu = \partial_\mu
- i \g B_\mu$. To have a well-defined decomposition into twist-two and three
contributions we extract tensors ${\cal R}_{\rho; \mu_1 \dots \mu_j}$ with
$(j + 1)$ indices corresponding to the two Young tables
\unitlength0.4cm
\begin{picture}(9,1)
\linethickness{0.06mm}
\put(1,0){\framebox(1,1){$\scriptstyle \rho$}}
\put(2,0){\framebox(1,1){$\scriptstyle \mu_1$}}
\put(3,0){\framebox(1,1){$\scriptstyle \mu_2$}}
\put(4,0){\framebox(3,1){$\ldots$}}
\put(7,0){\framebox(1,1){$\scriptstyle \mu_n$}}
\end{picture}
and
\unitlength0.4cm
\begin{picture}(8,1)
\linethickness{0.06mm}
\put(1,-1){\framebox(1,1){$\scriptstyle \rho$}}
\put(1,0){\framebox(1,1){$\scriptstyle \mu_1$}}
\put(2,0){\framebox(1,1){$\scriptstyle \mu_2$}}
\put(3,0){\framebox(3,1){$\ldots$}}
\put(6,0){\framebox(1,1){$\scriptstyle \mu_n$}}
\put(7.3,0){.}
\end{picture}
This can easily be done with the result
\begin{eqnarray}
\label{NonLoc-Loc-Ope}
{\cal O}_\rho (\kappa, - \kappa)
&=& {\cal R}^2_\rho (\kappa, - \kappa)
+ {\cal R}^3_\rho (\kappa , - \kappa)
\nonumber\\
&=& \sum_{j = 0}^{\infty} \frac{(-i\kappa)^j}{j!}
n_{\mu_1} \dots n_{\mu_j}
\left\{ {\cal R}^2_{\rho; \mu_1 \dots \mu_j}
+ \frac{2 j}{j + 1} {\cal R}^3_{\rho; \mu_1 \dots \mu_j}
\right\} ,
\end{eqnarray}
with
\begin{equation}
{\cal R}^2_{\rho; \mu_1 \dots \mu_j}
= \Sym_{\rho \mu_1 \dots \mu_j}
\bar\psi {\mit\Gamma}_\rho \,
i\!\stackrel{\leftrightarrow}{\cal D}_{\mu_1}
\dots
i\!\stackrel{\leftrightarrow}{\cal D}_{\mu_j}
\psi , \qquad
{\cal R}^3_{\rho; \mu_1 \dots \mu_j}
= \Sym_{\mu_1 \dots \mu_j} \Asym_{\rho \mu_1} \Sym_{\mu_1 \dots \mu_j}
\bar\psi {\mit\Gamma}_\rho \,
i\!\stackrel{\leftrightarrow}{\cal D}_{\mu_1}
\dots
i\!\stackrel{\leftrightarrow}{\cal D}_{\mu_j}
\psi ,
\end{equation}
where $\displaystyle{ \tSym_{\mu_1 \dots \mu_j}}$ is a symmetrization
(and trace subtraction) operation of $j$ indices with weight
$\frac{1}{j!}$ and antisymmetrization being defined by ${\displaystyle
\tAsym_{\mu_1 \mu_2}} t_{\mu_1\mu_2} = \ft12 \left( t_{\mu_1\mu_2} -
t_{\mu_2\mu_1} \right)$. In terms of light-ray operators the twist-two
part reads
\begin{eqnarray}
\label{Tw3operator}
{\cal R}^2_\rho (\kappa, - \kappa)
&=& \int_{0}^{1} du
\Bigg\{
\bar\psi (- u \kappa n) {\mit\Gamma}_\rho \psi (\kappa n) \nonumber\\
&+& \frac{\kappa}{2}
\int_{- u}^{u} d \tau
\bar\psi (- u \kappa n) {\mit\Gamma}_+
\left[
\stackrel{\rightarrow}{\partial}_{\rho} (u \kappa n)
-
\stackrel{\leftarrow}{\partial}_{\rho} (- u \kappa n)
- 2 i \g B_\rho (\tau \kappa n)
\right]
\psi (u \kappa n)
\Bigg\}
\end{eqnarray}
in the light-cone gauge $B_+ = 0$. If we would work in a covariant
gauge this would result, apart from restoration of the gauge-link
factors, to the substitution in the square brackets
\begin{equation}
\stackrel{\rightarrow}{\partial}_{\rho} (u \kappa n)
-
\stackrel{\leftarrow}{\partial}_{\rho} (- u \kappa n)
- 2 i \g B_\rho (\tau \kappa n)
\ \to \
\stackrel{\rightarrow}{\cal D}_{\rho} (u \kappa n)
-
\stackrel{\leftarrow}{\cal D}_{\rho} (- u \kappa n)
+ 2 i \, \tau \kappa \, \g G_{+ \rho} (\tau \kappa n) .
\end{equation}
One can easily project onto the twist-two part from the operator ${\cal
O}_\rho$ by contraction with the vector $n_\rho$, namely,
${\cal R}^2_+ = {\cal O}_+$.

The conversion of the twist-three part in terms of three-particle
operators is more tricky as compared to the forward case due to 
relevance of operators involving total derivatives. From the Eqs.\ 
(\ref{NonLoc-Loc-Ope}) and (\ref{Tw3operator}) one can easily derive, 
however, the following equation, cf.\ \cite{BalBra89},
\begin{eqnarray}
\label{recursion}
{^V\!\!{\cal O}_\rho} (\kappa, - \kappa)
&=& {^V\!{\cal R}_\rho^2} (\kappa, - \kappa)
+ \kappa \int_{0}^{1} du
\Bigg\{
i u \, \epsilon_{+ \rho \mu \nu} \partial_\mu
{^A{\cal O}_\nu} (u \kappa, - u \kappa) \nonumber\\
&&+ \frac{\kappa}{2} \int_{- u}^{u} d\tau
\left[
(u - \tau)
{^V\!\!{\cal S}^{+}_\rho} (u \kappa, \tau \kappa, - u \kappa)
-
(u + \tau)
{^V\!\!{\cal S}^{-}_\rho} (u \kappa, \tau \kappa, - u \kappa)
\right]
\Bigg\} ,
\end{eqnarray}
with $\partial_\mu = \stackrel{\rightarrow}{\partial}_\mu +
\stackrel{\leftarrow}{\partial}_\mu$ being the total derivative, and
similar expression for ${^A{\cal O}_\rho}$ which is obtained upon
substitution ${\mit\Gamma}_\rho \to {\mit\Gamma}_\rho \gamma_5$. These
two equations set an infinite iteration in total derivatives. We can
represent these relations in a schematic form
\begin{eqnarray*}
{^V\!\!{\cal O}} = {^V\!{\cal R}^2}
+ {\cal J} \ast
\left( \epsilon \, \partial \, {^A{\cal O}} + {^V\!\!{\cal S}} \right) ,
\qquad
{^A{\cal O}} = {^A{\cal R}^2}
+ {\cal J} \ast
\left( \epsilon \, \partial \, {^V\!\!{\cal O}} + {^A\!{\cal S}} \right) ,
\end{eqnarray*}
with ${\cal J}$ being an integral operator. A formal solution to this
system of equations reads
\begin{eqnarray*}
\label{Translation}
{^V\!\!{\cal O}} = \left( 1 - {\cal J}^2 \partial^2 \right)^{- 1}
\ast
\left\{ {^V\!{\cal R}^2}
+ \epsilon \, \partial \, {\cal J} \ast {^A{\cal R}^2}
+ {\cal J} \ast \left( {\cal J} \partial + 1 \right) \ast
{^V\!\!{\cal S}} \right\} ,
\end{eqnarray*}
and similar for ${^A{\cal O}}$. So that ${^I\!{\cal O}} - {^I\!{\cal R}^2}$
is a twist-three part of ${^I{\cal O}}$. The series can be summed up
into total translations in the light cone formalism. Let us demonstrate
first the effect of total derivatives in the basis of local operators. The
moments of Eq.\ (\ref{recursion}) look like
\begin{equation}
{^V\!\!{\cal O}_{\rho; j}} = {^V\!{\cal R}^2_{\rho; j}}
+ \frac{j}{j + 1} \, i \epsilon_{+ \rho \mu \nu} \, i \partial_\mu \,
{^A{\cal O}_{\nu; j - 1}}
-\frac{2}{j + 1} \sum_{k = 1}^{j - 1}
\left\{ ( j - k ) {^V\!\!{\cal S}^+_{\rho; j, k}}
- k {^V\!\!{\cal S}^-_{\rho; j, k}} \right\} ,
\end{equation}
where obviously ${\cal O}_{\rho; j} = n_{\mu_1} \dots n_{\mu_j}
{\cal O}_{\rho; \mu_1 \dots \mu_j}$ and we have introduced
three-particle local operators
\begin{equation}
{^V\!\!{\cal S}^{\pm}_{\rho; j, k}}
= i \g \, i^{j - 2}\bar\psi
\stackrel{\leftrightarrow}{\partial}\!\!{}^{k - 1}_+
\left[
\gamma_+ G_{+ \rho} \pm i \gamma_+ \gamma_5 {\widetilde G}_{+ \rho}
\right]
\stackrel{\leftrightarrow}{\partial}\!\!{}^{j - k - 1}_+
\psi ,
\end{equation}
and the analogous one for odd parity, i.e.\ $V \rightarrow A$ and 
$\gamma_+ \to \gamma_+ \gamma_5$. The solution
for ${\cal R}^3$ reads
\begin{eqnarray}
\label{Sol-R3}
{^V\!{\cal R}^3_{\rho; j}}
\!\!\! &=&\!\!\! \frac{1}{2 j}
\sum_{l = 0}^{j - 1} (j - l) \, (i\partial_+)^l
\left\{
\sigma_{l + 1} \, i \epsilon_{+ \rho \mu \nu} \, i\partial_\mu \,
{^A{\cal R}^2_{\nu; j - l - 1}}
-
\sigma_l \,
\left( i\partial_\rho n_\sigma - g_{\rho\sigma} i\partial_+ \right) \,
{^V\!{\cal R}^2_{\sigma; j - l - 1}}
\right\} \nonumber\\
&-&\!\!\! \frac{1}{j}
\sum_{l = 0}^{j - 2} \sum_{k = 1}^{j - l - 1}
(i\partial_+)^l
\left\{
( j - k - l ) {^V\!\!{\cal S}^{+}_{\rho; j - l, k}}
- (-1)^l k \, {^V\!\!{\cal S}^{-}_{\rho; j - l, k}}
\right\} ,
\end{eqnarray}
with $\sigma_l = \frac{1}{2} [1 - (- 1)^l]$. Now we can resum this equation
back into light-ray operators, or this can directly be obtained from Eq.\
(\ref{Translation}), with the result
\begin{eqnarray}
\label{R3-part}
{^V\!{\cal R}^3_\rho} (\kappa , - \kappa)
\!\!\!&=&\!\!\! \frac{\kappa}{2} \int_{0}^{1} du
\Bigg\{
u \, i \epsilon_{+ \rho \mu \nu} \partial_\mu
\left[
{^A{\cal R}^2_\nu} \left( \kappa, (\bar u - u) \kappa \right)
+
{^A{\cal R}^2_\nu} \left( (u - \bar u) \kappa , - \kappa \right)
\right] \\
&&\qquad\quad- u \,
\left( \partial_\rho n_\sigma - g_{\rho\sigma} \partial_+ \right) \,
\left[
{^V\!{\cal R}^2_\sigma} \left( \kappa, (\bar u - u) \kappa \right)
-
{^V\!{\cal R}^2_\sigma} \left( (u - \bar u) \kappa , - \kappa \right)
\right] \nonumber\\
+\kappa\!\!\!\!\!\!\!\!\!&&\!\!\!\!\int_{- u}^{u} d \tau
\left[
(u - \tau) {^V\!\!{\cal S}^{+}_\rho}
\left( \kappa, (\tau + \bar u) \kappa, (\bar u - u) \kappa \right)
-
(u + \tau) {^V\!\!{\cal S}^{-}_\rho}
\left( (u - \bar u) \kappa, (\tau - \bar u) \kappa, - \kappa \right)
\right]
\Bigg\} , \nonumber
\end{eqnarray}
and same for $V \leftrightarrow A$. Contrary to the forward scattering
there appeared contributions from different parity operators to a given
twist-three one and the `center-of-mass' of two- and three-particle
operators gets shifted by a total translation $\exp \left( \pm i \bar u
\kappa \partial \right)$. This equation gives a relation between skewed
parton distributions of different `twists' when sandwiched between
hadronic states, which we discuss in the next section.

%%%%%%%%%%%%%%%%%%%%%%%%%%%%%%%%%%%%%%%%%%%%%%%%%%%%%%%%%%%%%%%%%%%%%
\section{Twist-three skewed parton distributions.}
%%%%%%%%%%%%%%%%%%%%%%%%%%%%%%%%%%%%%%%%%%%%%%%%%%%%%%%%%%%%%%%%%%%%%

In this section we define twist-three skewed parton distributions
and their relation to twist-two ones. Before we deal with spin-$\ft12$
functions, we consider first a more simple case of spinless target.
For generality, we deal with the incoming and outgoing hadrons of
different masses $P_1^2 = M_1^2 \neq P_2^2 = M_2^2$.

%%%%%%%%%%%%%%%%%%%%%%%%%%%%%%%%%%%%%%%%%%%%%%%%%%%%%%%%%%%%%%%%%%%%%
\subsection{Spin-0 target.}
\label{Spin0}
%%%%%%%%%%%%%%%%%%%%%%%%%%%%%%%%%%%%%%%%%%%%%%%%%%%%%%%%%%%%%%%%%%%%%

It is instructive to start the analysis with the expectation values of
local operators. Since for spin-0 hadron we can not form a `twist-two'
axial-vector, only the vector twist-two operator develops non-zero
reduced matrix elements, which are given by
\begin{eqnarray}
\label{LorentzVscal}
\langle P_2| {^V\!{\cal R}^2_{\rho; \mu_1 \dots \mu_j}} |P_1\rangle
\!\!\!&=&\!\!\! \Sym_{\rho \mu_1 \dots \mu_j}
\left\{
P_\rho \dots P_{\mu_j} B_{j + 1, j + 1}
+ \Delta_\rho P_{\mu_{1}} \cdots P_{\mu_{j}} B_{j + 1, j}
+ \cdots
+ \Delta_\rho\dots \Delta_{\mu_j} B_{j + 1, 0}
\right\}, \nonumber\\
\langle P_2| {^A{\cal R}^2_{\rho; \mu_1 \dots \mu_j}} |P_1\rangle
\!\!\!&=&\!\!\! 0.
\end{eqnarray}
The reduced matrix elements $B_{jk}$ are defined as moments of a
skewed parton distribution $B (x, \eta)$:
\begin{eqnarray}
\label{Def-SPD-mom}
B_{j, j - k} = \frac{1}{k!}
\frac{d^k}{d \eta^k} \int_{-1}^{1} dx\; x^{j - 1}
B(x,\eta)_{|\eta = 0},
\qquad\mbox{where}\qquad
0 \le k \le j, \quad 1 \le j.
\end{eqnarray}
For the analysis of the moments of skewed parton distributions we
have to project both sides of (\ref{LorentzVscal}) with the light-cone
vectors $n_{\mu_1}\dots n_{\mu_j}$. Since
\begin{eqnarray}
n_{\mu_1} \dots n_{\mu_j} \Sym_{\rho \mu_1 \dots \mu_j}
\Delta_\rho \dots \Delta_{\mu_k} P_{\mu_{k+1}} \dots P_{\mu_j}
\!\!\!&=&\!\!\!
\frac{1}{j + 1}
\left\{
k \Delta_\rho \Delta_+^{k-1} P_+^{j + 1 - i}
+ (j + 1 - k) P_\rho \Delta_+^{k} P_+^{j-k}
\right\}
\nonumber\\
&=&\!\!\! \frac{1}{j + 1}
\left\{
k \Delta_\rho \eta^{k-1} + (j + 1 - k) P_\rho \eta^{k}
\right\} P_+^{j},
\end{eqnarray}
the coefficients in front of the two monomials are generated by
derivatives of $B_{j + 1}(\eta)$ w.r.t.\ the skewedness parameter, namely,
\begin{eqnarray}
\label{Def-Tw2-Sca}
&&\langle P_2| {^V\!{\cal R}^2_{\rho;j}} |P_1 \rangle
= P_{\rho} P_+^{j}
\left( 1 - \frac{\eta}{j+1} \frac{d}{d\eta} \right) B_{j + 1} (\eta)
+ \Delta_{\rho} P_+^{j} \frac{1}{j + 1} \frac{d}{d\eta} B_{j + 1} (\eta),
\nonumber\\
&&\langle P_2| {^V\!{\cal R}^2_{\rho;j}} |P_1 \rangle n_\rho
= P_+^{j + 1} \ B_{j + 1} (\eta),
\end{eqnarray}
with
\begin{eqnarray*}
B_{j + 1} (\eta) = \sum_{k = 0}^{j+1} \eta^k \ B_{j + 1, j + 1 - k}
= \int_{-1}^{1} dx\; x^j B (x, \eta) ,
\end{eqnarray*}
The moments $B_{j + 1} (\eta)$ can be taken from the $+$ component
of the operator ${\cal R}_+$. As a consequence of symmetrization the
first equation contains a Wandzura-Wilczek term proportional to
$\Delta_{\rho}^\perp = \Delta_\rho - \eta P_\rho$, which effectively
enters as a twist-three contribution to the scattering amplitude. The
matrix element of the light-ray operator can be obtained in a
straightforward manner by resummation:
\begin{eqnarray}
\langle P_2 | {^V\!{\cal R}^2_\rho} (\kappa, - \kappa)| P_1\rangle
= \int_{-1}^1 dx e^{-i \kappa P_+ x}
\left(
P_\rho B (x, \eta)
+ \Delta_\rho^\perp \int_{-1}^{1} dy \,
W_2 (x, y) \frac{d}{d\eta} B (x, \eta)
\right),
\end{eqnarray}
where the kernel reads $W_2(x, y) = \theta(x) \theta(y - x)/y
+ \theta(-x) \theta(x - y)/(-x)$.

Next we define the reduced matrix elements of the antiquark-gluon-quark
operators. Since these operators are partially antisymmetrized, we have
obviously two vectors $\Delta_{\rho}^\perp$ and the dual ones
$\widetilde\Delta_{\rho}^\perp = i \epsilon^\perp_{\rho \sigma}
\Delta^\perp_\sigma$, introduced above, in our disposal. Thus, the
general decomposition of reduced matrix elements reads
\begin{eqnarray}
\label{Def-SV-Loc-Red}
\langle P_2| {^V\!\!{\cal S}^{\pm}_{\rho; j, k}} |P_1\rangle
= \Delta_{\rho}^\perp S^{\pm}_{j, k} P_+^j ,
\qquad
\langle P_2| {^A\!{\cal S}^{\pm}_{\rho; j, k}} |P_1\rangle
= \widetilde \Delta_{\rho}^\perp \widetilde R^{\pm}_{j, k} P_+^j.
\end{eqnarray}
The duality relation (\ref{Pro-Dual}) reduces immediately the number
of independent contributions to two instead of four, namely,
$\widetilde R^{\pm}_{j, k} =  S^{\pm}_{j, k}$.  It turns out convenient
to work in a mixed representation for the skewed parton distributions.
We introduce a representation that depends on the position of the gluon
field and a Fourier conjugate variable with respect to a distance
between both quark fields:
\begin{eqnarray}
\langle P_2|\left\{
{{^V\!\!{\cal S}^{\pm}_{\rho}} ( \kappa, u \kappa, - \kappa )
\atop {^A\!{\cal S}^{\pm}_{\rho}} ( \kappa, u \kappa, - \kappa) }
\right\}|P_1 \rangle
\!\!\!&=&\!\!\! P_+^2
\left\{{\Delta_{\rho}^\perp \atop \widetilde\Delta_{\rho}^\perp } \right\}
 \int dx e^{-i \kappa x P_+}
S^{\pm}( x, u, \eta ).
\end{eqnarray}
The moments with respect to the momentum fraction $x$ are given by
a polynomial of order $j - 2$ in the variable $u$:
\begin{eqnarray}
S^{\pm}_j (u, \eta) \equiv \int_{-1}^1 dx \, x^{j-2} S^{\pm} (x, u, \eta)
= \sum_{k = 1}^{j - 1} \left( {j - 2 \atop k - 1 } \right)
\left( \frac{1 + u}{2} \right)^{k - 1}
\left( \frac{1 - u}{2} \right)^{j - k - 1} S^{\pm}_{j, k} (\eta),
\end{eqnarray}
where $S_{j, k}(\eta)$ are polynomials in $\eta$ of order $j$ defined
in Eq.\ (\ref{Def-SV-Loc-Red}).  As a simple consequence of our definition
we have
\begin{eqnarray}
\label{Int-SV-Ope}
\int_{-1}^1 du \frac{1 + u}{2} S^{\pm}_j (u, \eta)
= 2\sum_{k = 1}^{j - 1} \frac{k}{j (j - 1)} S^{\pm}_{j,k} (\eta),
\quad
\int_{-1}^1 du \frac{1 - u}{2} S^{\pm}_j (u, \eta)
= 2\sum_{k = 1}^{j - 1} \frac{j - k}{j (j - 1)} S^{\pm}_{j,k} (\eta) .
\end{eqnarray}
Finally, to find the matrix element of the operator ${\cal O}_{\rho; j}
= {\cal R}^2_{\rho; j} + \frac{2 j}{j + 1} {\cal R}^3_{\rho; j}$, it
remains to insert our findings (\ref{Def-Tw2-Sca}) and (\ref{Int-SV-Ope})
into the solution (\ref{Sol-R3}):
\begin{eqnarray}
P_+^{- j} \langle P_2| {^V\!\!{\cal O}_{\rho; j}} |P_1\rangle
\!\!\!&=&\!\!\! \left( P_{\rho} + \frac{\Delta_{\rho}^\perp}{\eta} \right)
B_{j + 1} (\eta)
+ \Delta_{\rho}^\perp
\left\{
\sum_{k = 0}^{j} \frac{\sigma_{j + 1 - k}}{j + 1}(-\eta)^{j - k}
\left( \frac{d}{d\eta} - \frac{k + 1}{\eta} \right) B_{k + 1}(\eta)
\right\}
\nonumber\\
&-&\!\!\!
\Delta_{\rho}^\perp
\sum_{k = 2}^{j} \frac{k! (-\eta)^{j - k}}{(j + 1)(k - 2)!}
\int_{-1}^1 du \left\{ \frac{1 - u}{2} S^{+}_{k} (u, \eta)
- (-1)^{j - k} \frac{1 + u}{2} S^{-}_{k} (u, \eta) \right\}.
\nonumber\\
\end{eqnarray}

Now we consider the axial-vector case, which is handled in the same way.
The main difference is that there is no twist-two part for spinless target.
However, a non-vanishing twist-two part of the vector operator induces now
a Wandzura-Wilczek type relation due to the epsilon tensor in Eq.\
(\ref{Sol-R3}) and provides
\begin{eqnarray}
P_+^{- j} \langle P_2| {^A{\cal O}_{\rho; j}} |P_1 \rangle
\!\!\!&=&\!\!\! \widetilde\Delta_{\rho}^\perp
\sum_{k = 0}^{j} \frac{\sigma_{j - k}}{j + 1} (-\eta)^{j - k}
\, \left( \frac{d}{d\eta} - \frac{k + 1}{\eta} \right) B_{k + 1}(\eta) \\
&-&\!\!\!
\widetilde\Delta_{\rho}^\perp \sum_{k = 2}^{j}
\frac{k! (-\eta)^{j - k}}{(j + 1)(k - 2)!}
\int_{-1}^1 du \left\{ \frac{1 - u}{2} R^{+}_{k} (u, \eta) -
(-1)^{j - k} \frac{1 + u}{2} R^{-}_{k} (u, \eta) \right\} .
\nonumber
\end{eqnarray}

The final step is a summation of local operators, see Eq.\
(\ref{NonLoc-Loc-Ope}), which leads to the expectation values of the
light-ray operators in terms of Fourier transform of skewed parton
distributions in parity even and odd cases
\begin{eqnarray}
\label{Res-Spi-O}
&&\langle P_2| {^V\!\!{\cal O}_{\rho}} (\kappa, -\kappa) |P_1 \rangle
= \int_{-1}^1 dx e^{-i \kappa P_+ x}
\Bigg\{ \Delta_\rho^\perp  \int_{-1}^{1}\frac{dy}{|\eta|}
W_+ \! \left( \frac{x}{\eta}, \frac{y}{\eta} \right)
\left(
\stackrel{\rightarrow}{\frac{d}{d\eta}}
- \frac{y}{\eta} \stackrel{\leftarrow}{\frac{d}{dy}}
\right) B(y,\eta) \\
&&\qquad\qquad\qquad\qquad\ \
+ \left(P_\rho + \frac{1}{\eta} \Delta^\perp_\rho \right) B(x,\eta)
\nonumber\\
&&\qquad - \Delta_{\rho}^\perp \int_{-1}^1 \frac{dy}{|\eta|} \int_{-1}^1 du
\left[ \frac{1 - u}{2}
W^{\prime\prime} \! \left( \frac{x}{\eta}, \frac{y}{\eta} \right)
S^+ (y, u, \eta)
+ \frac{1 + u}{2}
W^{\prime\prime} \! \left( - \frac{x}{\eta}, - \frac{y}{\eta} \right)
S^- (y, u, \eta) \right]  \Bigg\}, \nonumber\\
&&\langle P_2| {^A{\cal O}_{\rho}} (\kappa,-\kappa) |P_1 \rangle
= \int_{-1}^1 dx e^{-i \kappa P_+ x} \Bigg\{
\widetilde\Delta_\rho^\perp
\int_{-1}^{1} \frac{dy}{|\eta|}
W_-\! \left( \frac{x}{\eta}, \frac{y}{\eta} \right)
\left(
\stackrel{\rightarrow}{\frac{d}{d\eta}}
- \frac{y}{\eta} \stackrel{\leftarrow}{\frac{d}{dy}}
\right) B(y, \eta) \\
&&\qquad - \widetilde \Delta_{\rho}^\perp
\int_{-1}^1 \frac{dy}{|\eta|} \int_{-1}^1 du
\left[ \frac{1 - u}{2}
W^{\prime\prime} \! \left( \frac{x}{\eta},\frac{y}{\eta} \right)
R^+ (y, u, \eta)
+ \frac{1 + u}{2}
W^{\prime\prime} \! \left( - \frac{x}{\eta}, - \frac{y}{\eta} \right)
R^- (y, u, \eta) \right] \Bigg\} .
\nonumber
\end{eqnarray}
Here $W^{\prime\prime} \left( \pm \frac{x}{\eta}, \pm \frac{y}{\eta} \right)
\equiv \frac{d^2}{dy^2} W\! \left( \pm \frac{x}{\eta}, \pm \frac{y}{\eta}
\right)$ and the $W$ kernels read
\begin{eqnarray}
&&W (x, y) = \frac{{\mit\Theta} (x, y)}{1 + y} ,
\quad\mbox{with}\quad
{\mit\Theta} (x, y) = {\rm sign} (1 + y)
\theta\left( \frac{1 + x}{1 + y} \right)
\theta\left(\frac{y - x}{1 + y}\right) , \\
&&W_\pm (x, y) = \frac{1}{2} \left\{ W (x, y) \pm W(-x, -y) \right\} .
\nonumber
\end{eqnarray}
In restoration of non-local form one uses the following result for
Mellin moments of $W$-kernels
\begin{equation}
\int_{-1}^1 \frac{dx}{|\eta|} x^j
W\! \left( \frac{x}{\eta}, \frac{y}{\eta} \right)
= \sum_{k = 0}^{j} \frac{(-1)^{j - k}}{j + 1} \eta^{j - k} y^k,
\qquad
\int_{-1}^1 \frac{dx}{|\eta|} x^j
W_\pm \! \left( \frac{x}{\eta}, \frac{y}{\eta} \right)
= \sum_{k = 0}^{j} \frac{(-1)^{j - k} \pm 1}{2(j + 1)} \eta^{j - k} y^k .
\end{equation}
Note that the piece $\frac{1}{\eta} \Delta^\perp_\rho B(x,\eta)$
in Eq.\ (\ref{Res-Spi-O}) is cancelled by a term arising from the convolution
with the $W_+$ kernel.

%%%%%%%%%%%%%%%%%%%%%%%%%%%%%%%%%%%%%%%%%%%%%%%%%%%%%%%%%%%%%%%%%%%%%
\subsection{Spin-$\ft12$ target.}
%%%%%%%%%%%%%%%%%%%%%%%%%%%%%%%%%%%%%%%%%%%%%%%%%%%%%%%%%%%%%%%%%%%%%

Now we are in a position to discuss a spin-$\ft12$ target. It is
convenient to express the expectation value of local operators in
terms of spinor bilinears
\begin{eqnarray}
\label{Def-ForFac}
&&\left( \, b, \widetilde b \, \right)
= \bar U (P_2, S_2) \left( 1, \gamma_5 \right) U(P_1, S_1),
\quad\,
\left( h_{\rho}, \widetilde{h}_\rho \right)
= \bar U (P_2, S_2) \gamma_\rho \left( 1, \gamma_5 \right) U (P_1, S_1),
\nonumber\\
&&\left( t_{\rho\sigma}, \widetilde{t}_{\rho\sigma} \right)
= \bar U (P_2, S_2) i \sigma_{\rho\sigma}
\left( 1, \gamma_5 \right) U (P_1, S_1),
\end{eqnarray}
Obviously, the dual tensor form factor $\widetilde t_{\rho\sigma}$ is
obtained from $t_{\rho\sigma}$ by contraction with the $\epsilon$-tensor
and can, therefore, be eliminated. Furthermore, equation of motion shows
that in each parity sector we have the relation between the structures
(\ref{Def-ForFac})
\begin{eqnarray}
&& P_\rho b = M_+ h_\rho  - t_{\rho\sigma} \Delta^\sigma ,
\qquad
\Delta_\rho b = M_- h_\rho  - t_{\rho\sigma} P^\sigma ,
\\
&& \Delta_\rho \widetilde b
= M_+ \widetilde h_\rho  - \widetilde t_{\rho\sigma} P^\sigma ,
\qquad
P_\rho \widetilde b
= M_- \widetilde h_\rho  - \widetilde t_{\rho\sigma} \Delta^\sigma,
\nonumber
\end{eqnarray}
where $M_\pm = M_2 \pm M_1$. As demonstrated above for a scalar target,
the symmetrization provides us a Wandzura--Wilczek term proportional to
$\Delta^\perp$. To take advantage of the analysis already performed in
the preceding section and for a symmetrical handling of even and odd
parity sectors, our basis is spanned by the (pseudo) scalar and the
(axial) vector bilinears. At the end, we express the result in terms
of conventional ones, introduced by Ji \cite{Ji97},
\begin{eqnarray}
h_\rho,
\qquad
e_\rho = t_{\rho\sigma} \frac{\Delta^\sigma}{M_+},
\qquad
\widetilde h_\rho,
\qquad
\widetilde e_\rho = \frac{\Delta_\rho} {M_+} \widetilde b .
\end{eqnarray}
Obviously, for the (pseudo) scalar bilinears we can take the results
deduced for scalar target. The only new structure for the spin-$\ft12$
target is proportional to the $h_\rho$, $\widetilde{h}_\rho$.

In parallel to section \ref{Spin0}, the local matrix elements read
\begin{eqnarray}
\langle P_2| {^V\!{\cal R}^2_{\rho; \mu_1 \dots \mu_j}} |P_1\rangle
\!\!\!&=&\!\!\!
\Sym_{\rho \mu_1 \dots \mu_j}
h_\rho\left\{P_{\mu_{1}} \dots P_{\mu_j} A_{j + 1, j + 1}
+ \cdots + \Delta_{\mu_{1}} \dots \Delta_{\mu_j} A_{j + 1, 0}\right\}
\nonumber\\
&+&\!\!\! \frac{b}{M_+} \cdots ,
\end{eqnarray}
where the ellipses (here and later on) stand for the r.h.s.\ of Eq.\
(\ref{LorentzVscal}) (and corresponding equations from the preceding
subsection). Analogous relation holds for the parity odd case. The
moments of the skewed parton distribution are defined analogously to
Eq.\ (\ref{Def-SPD-mom}), $A_{j + 1}(\eta) = \int_{-1}^{1} dx\; x^j 
A(x, \eta)$. The difference arises only from the symmetrization procedure
\begin{eqnarray}
n_{\mu_1} \dots n_{\mu_j} \Sym_{\rho \mu_1 \dots \mu_j}
h_\rho \Delta_{\mu_1} \dots \Delta_{\mu_k} P_{\mu_{k + 1}} \dots P_{\mu_j}
= \frac{P_+^{j - 1}}{j + 1}\left\{ h_\rho \eta^{k} P_+
+ (j - k) P_\rho h_+ \eta^{k} +  k \Delta_\rho h_+ \eta^{k - 1} \right\} ,
\end{eqnarray}
and provides now a new structure
\begin{eqnarray}
\label{Def-Tw2-h}
&&\langle P_2| {^V\!{\cal R}^2_{\rho;j}} |P_1 \rangle
= \frac{P_+^{j - 1}}{j + 1}
\left( (j + 1) P_{\rho} h_+
+ h_{\rho} - P_{\rho} h_+ + \Delta_{\rho}^\perp h_+
\frac{d}{d\eta} \right) A_{j + 1}(\eta) + \frac{b}{M_+} \cdots ,
\nonumber\\
&& \langle P_2| {^V\!{\cal R}^2_{\rho;j}} |P_1 \rangle n_\rho
= h_+ P_+^j \ A_{j + 1}(\eta) 
+ \frac{b}{M_+} P_+^{j + 1} \ B_{j + 1}(\eta) .
\end{eqnarray}
Next we resum the local result and get
\begin{eqnarray}
&&\langle P_2 | {^V\!{\cal R}^2_\rho} (\kappa, - \kappa)| P_1 \rangle
= \int_{-1}^1 dx e^{-i \kappa P_+ x}
\Bigg\{
P_\rho \left( \frac{h_+}{P_+} A(x, \eta) + \frac{b}{M_+} B(x, \eta) \right)
\\
&&\qquad\qquad + \int_{-1}^{1} dy \, W_2(x,y)
\left[ \Delta_\rho^\perp
\left( \frac{h_+}{P_+} \frac{d}{d\eta} A(y, \eta)
+ \frac{b}{M_+} \frac{d}{d\eta} B (y, \eta) \right)
+ \left( h_{\rho} -P_{\rho} \frac{h_+}{P_+} \right)  A(y,\eta) \right]
\Bigg\} . \nonumber
\end{eqnarray}
Projecting this expression with vector $n_\rho$ we obtain conventional
definitions for twist-two skewed parton distributions. The basis used
presently can be easily expressed in terms of Ji's parametrization as
follows, for Dirac structures
\begin{equation}
\label{BilinearSubst}
\widetilde b = \frac{M_+}{\Delta_+} \widetilde e_+ ,
\qquad
b = \frac{M_+}{P_+} \left( h_+  - e_+ \right) ,
\end{equation}
and skewed parton distributions
\begin{eqnarray}
\label{SPDsSubst}
\widetilde A = \widetilde  H ,
\quad \widetilde B = \eta \widetilde E ,
\quad A = H + E ,
\quad B = - E .
\end{eqnarray}
For the following, it is useful to note as well that $M_+ \left( h_\rho -
\frac{P_\rho}{P_+} h_+ \right) = \left( t_{\rho\sigma} - \frac{P_\rho}{P_+}
t_{+\sigma} \right) \Delta_\sigma$.

For the time being we introduce three-particle skewed parton distributions
without an explicit spinor bilinear decomposition,
\begin{eqnarray}
\langle P_2| \left\{
{
{^V\!\!{\cal S}^{\pm}_{\rho}} (\kappa, u \kappa, - \kappa)
\atop
{^A\!{\cal S}^{\pm}_{\rho}}(\kappa, u \kappa, - \kappa)
}
\right\} |P_1 \rangle
=
P_+^2 \int dx e^{- i \kappa x P_+}
\left\{
{
S_\rho^{\pm}(x, u, \eta)
\atop
R_\rho^{\pm}(x, u, \eta)
}
\right\}.
\end{eqnarray}
A discussion of the parametrization will be given below.

Due to length of the consequent formulas we give below only results
for the vector case since the axial one follows from the former by
substitutions. Combining the Eq.\ (\ref{Def-Tw2-h}) with the
two-particle part of Eq.\ (\ref{Sol-R3}) we get for the $h$-part of
the local operators ${^V\!\!{\cal O}_{\rho}}$
\begin{eqnarray}
P_+^{1 - j} \langle P_2| {^V\!\!{\cal O}_{\rho;j}} |P_1 \rangle
\!\!\!&=&\!\!\!
\left( P_{\rho} + \frac{\Delta_{\rho}^\perp}{\eta} \right)
h_+ A_{j+1}
+ \sum_{k = 0}^{j} \frac{\sigma_{j + 1 - k}}{j + 1}(- \eta)^{j - k}
\Bigg\{
\Delta_{\rho}^\perp h_+ \left( \frac{d}{d\eta} - \frac{k + 1}{\eta} \right)
\\
&+& \left( h_\rho P_+ - P_\rho h_+ \right) \Bigg\} A_{k + 1}
+ \sum_{k = 0}^{j} \frac{\sigma_{j - k}}{j + 1}(- \eta)^{j - k}
\Bigg\{
\widetilde \Delta_{\rho}^\perp \widetilde h_+
\left( \frac{d}{d\eta} - \frac{k + 1}{\eta} \right) \nonumber\\
&+& i \epsilon^\perp_{\rho \sigma}
\left( \widetilde h_\sigma P_+ - P_\sigma \widetilde h_+ \right)
\Bigg\}  \widetilde A_{k+1} + \cdots . \nonumber
\end{eqnarray}
Transforming it finally to the non-local form, and adding the $b$-term and
the missing three-particle piece, one finds
\begin{eqnarray}
\label{ResFin-1/2}
&&\langle P_2| {^V\!\!{\cal O}_{\rho}} (-\kappa, \kappa) |P_1 \rangle
= \int_{-1}^1 dx e^{- i \kappa P_+ x}
\Bigg\{
\left( P_\rho + \frac{1}{\eta} \Delta^\perp_\rho \right) D (x, \eta) \\
&&\qquad\qquad+ \int_{-1}^{1} \frac{d y}{|\eta|}
\Bigg[ W_+\!\left( \frac{x}{\eta}, \frac{y}{\eta} \right)
\left(
\Delta_\rho^\perp
\left(
\stackrel{\rightarrow}{\frac{d}{d\eta}}
- \frac{y}{\eta} \stackrel{\leftarrow}{\frac{d}{dy}}
\right) D (y, \eta)
+ \left( h_\rho - P_\rho \frac{h_+}{P_+} \right) A (y, \eta) \right)
\nonumber\\
&&\qquad\qquad\qquad\quad\ \
+ W_- \! \left( \frac{x}{\eta}, \frac{y}{\eta} \right)
\left( \widetilde \Delta_\rho^\perp
\left(
\stackrel{\rightarrow}{\frac{d}{d\eta}}
- \frac{y}{\eta} \stackrel{\leftarrow}{\frac{d}{dy}}
\right) \widetilde{D} (y, \eta)
+ i \epsilon^\perp_{\rho\sigma}
\left( \widetilde h_\sigma - P_\sigma \frac{\widetilde h_+}{P_+} \right)
\widetilde A (y, \eta) \right) \nonumber\\
&&\qquad\qquad- \int_{-1}^1 du \left(
\frac{1 - u}{2}
W^{\prime\prime} \! \left( \frac{x}{\eta}, \frac{y}{\eta} \right)
S_\rho^+ (y, u, \eta)
+ \frac{1 + u}{2}
W^{\prime\prime} \! \left( - \frac{x}{\eta}, - \frac{y}{\eta} \right)
S_\rho^- (y, u, \eta) \right) \Bigg]  \Bigg\},
\nonumber
\end{eqnarray}
where we have introduced a shorthand notation for the
combinations
\begin{equation}
D (x, \eta) = \frac{h_+}{P_+} A (x, \eta) + \frac{b}{M_+} B (x, \eta) ,
\qquad
\widetilde{D} (y, \eta) = \frac{\widetilde h_+}{P_+} \widetilde A (y, \eta)
+ \frac{\widetilde b}{M_+} \widetilde B (y, \eta) .
\end{equation}
The derivatives w.r.t.\ $\eta$ in Eq.\ (\ref{ResFin-1/2}) acts {\it only}
on skewed parton distributions {\em but not} on spinor bilinears.
The axial-vector case is deduced by the following trivial substitutions:
$b \leftrightarrow \widetilde b, h_\rho \leftrightarrow \widetilde h_\rho,
B \leftrightarrow \widetilde B, A \leftrightarrow \widetilde A$ and
$S^\pm_\rho\to R^\pm_\rho$. Conventional form is obtained by means of
substitutions (\ref{BilinearSubst},\ref{SPDsSubst}). Equation
(\ref{ResFin-1/2}) is our final result. Its transverse part gives
an expression for the `transverse' twist-three skewed functions in terms
of `known' leading twist and three particle functions.

Let us address briefly the parametrization of genuine twist-three part
of the correlation functions. An analysis shows that there exist four
independent twist-three spinor bilinears, which we can choose to be
\begin{equation}
\Delta_\rho^\perp \frac{b}{M_+},
\qquad
\Delta_\rho^\perp \frac{h_+}{P_+};
\qquad
\Delta_\rho^\perp \frac{\widetilde b}{M_+},
\qquad
\Delta_\rho^\perp \frac{\widetilde h_+}{P_+};
\end{equation}
while the dual ones read in the same sequence
\begin{equation}
\widetilde \Delta_\rho^\perp \frac{b}{M_+},
\qquad
\widetilde \Delta_{\rho}^\perp \frac{h_+}{P_+};
\qquad
\widetilde\Delta_\rho^\perp \frac{\widetilde b}{M_+},
\qquad
\widetilde\Delta_{\rho}^\perp  \frac{\widetilde h_+}{P_+}.
\end{equation}
Note that a fifth candidate, $M_+ \frac{t_{+\rho}}{P_+}$, and trace terms
proportional to $ M_+^2 n_\rho$ enter at twist-four level.
One can immediately see that bilinears used in Eq.\ (\ref{ResFin-1/2})
can be spanned by this basis via relations $h_{\rho} - P_{\rho}
\frac{h_+}{P_+} = - \frac{\eta}{1 - \eta^2} \Delta_\rho^\perp
\frac{h_+}{P_+} + \frac{1}{1 - \eta^2} \widetilde \Delta_\rho^\perp
\frac{\widetilde h_+}{P_+}$ and $\widetilde h_{\rho} - P_{\rho}
\frac{\widetilde h_+}{P_+} = - \frac{\eta}{1 - \eta^2} \Delta_\rho^\perp
\frac{\widetilde h_+}{P_+} + \frac{1}{1 - \eta^2} \widetilde
\Delta_\rho^\perp \frac{h_+}{P_+}$. The parametrization of correlation
functions $S$ and $R$ then reads
\begin{eqnarray}
S_\rho^{\pm} (x, u, \eta)
&=& \Delta_{\rho}^\perp \frac{b}{M_+}\ S_1^{\pm}
+ \Delta_\rho^\perp \frac{h_+}{P_+} S_2^{\pm}
+ \widetilde\Delta_{\rho}^\perp \frac{\widetilde b}{M_+} \widetilde S_1^{\pm}
+ \widetilde\Delta_\rho^\perp \frac{\widetilde h_+}{P_+} \widetilde S_2^{\pm}
,\nonumber\\
R_\rho^{\pm}(x,u,\eta)
&=& \Delta_{\rho}^\perp \frac{\widetilde b}{M_+} R_1^{\pm}
+ \Delta_\rho^\perp \frac{\widetilde h_+}{P_+} R_2^{\pm}
+ \widetilde\Delta_{\rho}^\perp \frac{b}{M_+} \widetilde R_1^{\pm}
+ \widetilde \Delta_\rho^\perp \frac{h_+}{P_+} \widetilde R_2^{\pm} ,
\end{eqnarray}
where duality (\ref{Pro-Dual}) requires $\widetilde S_i^\pm = R_i^\pm,\
\widetilde R_i^\pm = S_i^\pm$. To reexpress the parity even and odd
sector only in terms of vector and axial-vector form vectors, respectively,
one can equally use an alternative basis, where the independent
elements are
\begin{equation}
\Delta_\rho^\perp \frac{b}{M_+} ,
\qquad
h_{\rho} - P_{\rho} \frac{h_+}{P_+} ,
\qquad
\Delta_\rho^\perp \frac{\widetilde b}{M_+} ,
\qquad
\widetilde h_{\rho}  - P_{\rho} \frac{\widetilde h_+}{P_+} ,
\end{equation}
while the dual ones read in the same sequence
\begin{equation}
\widetilde \Delta_\rho^\perp \frac{b}{M_+} ,
\qquad
\frac{\Delta_{\rho} \widetilde h_+ - \widetilde h_{\rho} \Delta_+}{P_+} ,
\qquad
\widetilde \Delta_\rho^\perp \frac{\widetilde b}{M_+} ,
\qquad
\frac{\Delta_{\rho} h_+ - h_{\rho} \Delta_+}{P_+}.
\end{equation}

%%%%%%%%%%%%%%%%%%%%%%%%%%%%%%%%%%%%%%%%%%%%%%%%%%%%%%%%%%%%%%%%%%%%%
\section{Power suppression of twist-three effects in DVCS.}
\label{suppressionDVCS}
%%%%%%%%%%%%%%%%%%%%%%%%%%%%%%%%%%%%%%%%%%%%%%%%%%%%%%%%%%%%%%%%%%%%%

Now we point out the phenomenological consequences of our analysis for the
DVCS process, $e N \to e' N' \gamma$, with $\eta = - \xi$. Since DVCS
interferes with the Bethe-Heitler process, the possibility exists to get a
direct access to the skewed parton distributions via the interference term.
In leading twist-two approximation of the DVCS hadronic tensor the amplitude
squares for unpolarized or longitudinal polarized nucleon behave as:
\begin{eqnarray}
|T_{\rm DVCS}|^2 \propto \frac{1}{Q^2},
\qquad
T_{\rm BH} T_{\rm DVCS}^\star
\propto
\frac{1}{\sqrt{- \Delta^2 Q^2}}
\left( 1 -  \frac{\Delta^2_{\rm min}}{\Delta^2} \right)^{1/2} ,
\qquad
|T_{\rm BH}|^2 \propto \frac{1}{\Delta^2},
\end{eqnarray}
where $\Delta_{\rm min}^2 = -4 M^2 \xi^2/ (1 - \xi^2)$ follows from the
kinematical boundary on which $\Delta^\perp$ vanish. Consequently, the
interference terms and so also the charge and spin asymmetries vanish
(this is not the case for transversely polarized target). Since the 
leading $\Delta_\perp$ dependence could also arise from the twist-three 
contributions to the DVCS amplitude, one may argue that those enter
in the interference term without power suppression. Consequently, the whole
twist-two analyses would be spoiled. This possible complication has been
already studied in the past \cite{BelMulNieSch00}, unfortunately, it was 
not clearly emphasized that twist-three contributions are power suppressed.

Our aim is to show that this is indeed the case, while the complete
interference term will be published elsewhere \cite{BelKirMueSch00}. In our
analysis \cite{BelMulNieSch00} we used the following parametrization of
the hadronic scattering amplitude that arises from the operator product
expansion in the free fermion theory, where current conservation was
restored in calculations by means of projection operators:
\begin{eqnarray}
T_{\mu\nu} (q, P, \Delta)
\!\!\!&=&\!\!\!
- {\cal P}_{\mu\sigma} g_{\sigma\tau} {\cal P}_{\tau\nu}
\frac{q \cdot V_1}{P \cdot q}
+ \left( {\cal P}_{\mu\sigma} P_\sigma  {\cal P}_{\rho\nu}
+ {\cal P}_{\mu\rho}  P_\sigma {\cal P}_{\sigma\nu} \right)
\frac{V_{2 \, \rho}}{P \cdot q^2} \\
&&\!\!\!
- {\cal P}_{\mu\sigma} i\epsilon_{\sigma \tau q \rho} {\cal P}_{\tau\nu}
\frac{A_{1 \, \rho}}{P \cdot q}
- {\cal P}_{\mu\sigma} i\epsilon_{\sigma \tau P \rho} {\cal P}_{\tau\nu}
\frac{A_{2 \,\rho}}{P \cdot q} , \nonumber
\end{eqnarray}
where ${\cal P}_{\mu\nu}$ has been introduced in section \ref{Amplitude}.
Comparing with the results given in Eqs.\
(\ref{GaugeInvAmpl},\ref{TensorStructure}), we find that the form
factors are related to each other in the following manner
\begin{eqnarray}
&& V_{1 \, \mu}
= \int \frac{dx}{\xi} C^{(-)}(x,\xi) {^V\!\!O_\mu} (x, \eta),
\qquad
A_{1 \, \mu}
= \int \frac{dx}{\xi} C^{(+)}(x,\xi) {^A O_\mu} (x, \eta),
\qquad
A_{2 \, \mu} = 0, \\
&& V_{2 \, \mu} = \int \frac{dx}{\xi}
\left\{
\xi  C^{(-)}(x, \xi)
\left(
{^V\!\!O_\mu} (x, \eta) - \frac{1}{2} \frac{P_\mu}{P \cdot q} \
q \cdot {^V\!\!O} \, (x, \eta)
\right)
+ C^{(+)}(x, \xi)
\frac{i}{2} \frac{\epsilon_{\mu \nu \Delta q}}{P \cdot q} \
{^A O_\nu} (x, \eta)
\right\} . \nonumber
\end{eqnarray}
A straightforward calculation and power counting show that in the case
of DVCS kinematics, where $\eta = - \xi$, the contributions of $V_2$ and
$A_2$ are power suppressed in comparison to the known twist-two result:
\begin{eqnarray}
{\cal T}_{\rm BH} {\cal T}_{\rm DVCS}^\ast
\!\!\!&=&\!\!\! - \frac{2 - 2y + y^2}{1 - y} \frac{\xi}{\Delta^2 q^4}
\left\{ \left( k_\sigma -\frac{1}{y} q_\sigma \right)
\left( J_\sigma - \Delta_\sigma \frac{q \cdot J}{q^2} \right)
q \cdot V_1^\dagger
+ i \epsilon_{k q \Delta J} \frac{q \cdot A_1^\dagger}{q^2} \right\}
\nonumber\\
&&\!\!\! - \frac{\lambda(2 - y) y}{1 - y} \frac{\xi}{\Delta^2 q^4}
\left\{ \left( k^\sigma -\frac{1}{y} q^\sigma \right)
\left( J_\sigma - \Delta_\sigma \frac{q \cdot J}{q^2} \right)
q \cdot A_1^\dagger
+ i \epsilon_{k q \Delta J} \frac{q \cdot V_1^\dagger}{q^2} \right\} .
\end{eqnarray}
Here $J_\mu$ is the electromagnetic current, $\lambda$ and $k_\mu$ denote
the polarization and momentum of the incoming electron. We employed a
simple rule that $q_\mu$, $k_\mu - \frac{1}{y} q_\mu$, and $P_\mu,
\Delta_\mu$, when contracted with an electromagnetic current or spinor
bilinears, give terms of order $Q^2$, $Q^1$, and $Q^0$, respectively. Note
that in agreement with this counting our results for unpolarized nucleon
target coincide with Ref. \cite{DieGouPirRal97}, where the contribution
of $V_2$ has been taken into account. Let us remark, that also in the case
of a scalar target the same situation holds true, as it already has been
shown in \cite{AniPirTer00}.

Since, the twist-three contributions enter as $1/Q$ power suppressed term to
all single spin and charge asymmetries, the comparison of models for twist-2
skewed parton distributions with experimental data will be contaminated in
an expected manner. Moreover, we emphasize again that the twist-3 functions
are completely known in terms of twist-two ones in absence of the gluonic
contributions as it is the case in the naive parton model. It is an
interesting non-perturbative problem to estimate the size of the gluonic
contributions in comparison to the Wandzura-Wilczek term. Assuming that the
dynamical twist-three effects encoded in the three-particle operators are
small, which presumably happens in view of recent experimental data
\cite{SLAC00} and new lattice results \cite{Lattice} for the forward
kinematics, one can use the Wandzura-Wilczek part of the relation
(\ref{ResFin-1/2}) as a model for the `transverse' twist-three skewed
parton distributions, $\langle P_2 | {^I\!{\cal O}_\mu^\perp} | P_1 \rangle$.
This will allow to give a numerical estimate of the power suppressed
contributions to asymmetries for the kinematics of present experiments
\cite{BelKirMueSch00}.

%%%%%%%%%%%%%%%%%%%%%%%%%%%%%%%%%%%%%%%%%%%%%%%%%%%%%%%%%%%%%%%%%%%%%
\section{Logarithmic scaling violation.}
\label{Evolution}
%%%%%%%%%%%%%%%%%%%%%%%%%%%%%%%%%%%%%%%%%%%%%%%%%%%%%%%%%%%%%%%%%%%%%

Let us add a final remark on the logarithmic scaling violation of
twist-three functions. Obviously the Wandzura-Wilczek part evolves via
the familiar twist-two generalized exclusive evolution equation known
to two-loop accuracy nowadays \cite{BelMul98}. The three-particle piece
does not require a new study as well, since it was elaborated in great
detail recently in the context of twist-three functions measured in
inclusive reactions (forward limit restrictions was not made in those
studies). Namely, the operators ${\cal S}$ alluded to above fall into
a class of the so-called quasi-partonic ones \cite{BukFroLipKur85} and
thus at leading order in the coupling constant their evolution kernel
is given by a sum of conventional skewed twist-two ones with appropriate
quantum numbers in subchannels: $\mbox{\boldmath$K$}_{\bar q g q} =
K_{\bar q g} + K_{g q} + K_{\bar q q}$. The issue of its diagonalization
(see a recent review \cite{Bel00}) has been addressed in detail recently
\cite{BraDerMan98,Bel99,DerKorMan99} within the context of integrable open
spin chain models which arise in multicolour limit\footnote{The latter
gives a good approximation since the corrections to the leading result
are suppressed in $1/N_c^2$.}. Namely, for $N_c \to \infty$ the kernel
$K_{\bar q q}$ can be neglected
and, e.g.\ $\ft1{N_c} K_{\bar q g} = \psi ( \hat J_{\bar q g} + \ft32 )
+ \psi ( \hat J_{\bar q g} - \ft32 ) - 2 \psi (1)$ and $\ft1{N_c}
K_{g q} = \psi ( \hat J_{\bar q g} + \ft12 ) + \psi ( \hat J_{\bar q g}
- \ft12 ) - 2 \psi (1)$ for ${\cal S}^+$, and with $\ft32$ and $\ft12$
being interchanged for ${\cal S}^-$. Due to conformal invariance at tree
level, the kernels depend on the quadratic Casimir operator of the collinear
conformal group $SO (2,1)$ or, in other words, on the conformal spin in
the subchannel only, $\mbox{\boldmath$\hat J$}^2 = \hat J (\hat J - 1)$.
It turns out that antiquark-gluon-quark system admits an extra integral
of motion $Q_S$ and is thus completely integrable. The result of the
analysis allows to find the eigenfunctions ${\mit\Psi}$ and eigenvalues
${\cal E}$ of the system and thus solve the evolution equation,
\begin{eqnarray*}
S ( x_1, x_2, x_3 | Q^2 )
= \sum_{\{ \alpha \}}
{\mit\Psi}_{\{ \alpha \}} (x_1, x_2, x_3)
\left(
\frac{\alpha_s ( Q_0^2 )}{\alpha_s ( Q^2 )}
\right)^{N_c {\cal E}_{\{ \alpha \}} / \beta_0}
\langle\langle {\cal S}_{\{ \alpha \}} ( Q_0^2 ) \rangle\rangle ,
\qquad
\sum_{i = 1}^{3} x_i = \eta .
\end{eqnarray*}
Here $\beta_0 = \frac{4}{3} T_F N_f - \frac{11}{3} C_A$ is the QCD beta
function and a set of quantum numbers $\{ \alpha \}$ parametrizes solutions
and can be chosen as eigenvalues of the conformal spin $J$ and the charge
$Q_S$. Namely, the lowest anomalous dimension, ${\cal E}_0 (J)$, is known
exactly \cite{AliBraHil91}, while the rest of the spectrum, ${\cal E}_q (J)$,
can be described with a high accuracy using WKB approximation
\cite{BraDerMan98,Bel99,DerKorMan99}
with the results
\begin{eqnarray*}
{\cal E}_0 (J) = \psi (J + 3) + \psi (J + 4) - 2 \psi (1) - \frac{1}{2} ,
\quad
{\cal E}_Q (J) = 2 \ln J - 4 \psi (1)
+ 2 \, {\rm Re} \, \psi \left( \frac{3}{2} + i \eta_S \right)
- \frac{3}{2} ,
\end{eqnarray*}
where $\eta_S$ is related to the conserved charge by the relation
$\eta_S \equiv \ft12 \sqrt{2 Q_S/J^2 - 3}$ and obeys a WKB quantization
condition which, once solved, gives quantized values of ${\cal E}_Q$.

For practical purposes, one may also generalize the polynomial
reconstruction method as it was applied for the forward kinematics in Ref.\
\cite{GeyMueRob96bBelMue97Mue97} or use a brute force numerical
integration to solve the evolution equations.

%%%%%%%%%%%%%%%%%%%%%%%%%%%%%%%%%%%%%%%%%%%%%%%%%%%%%%%%%%%%%%%%%%%%%
\section{Conclusions.}
%%%%%%%%%%%%%%%%%%%%%%%%%%%%%%%%%%%%%%%%%%%%%%%%%%%%%%%%%%%%%%%%%%%%%

In this paper we have studied twist-three effects for the two-photon
amplitudes in the generalized Bjorken kinematics. We have
demonstrated explicitly the restoration of the electromagnetic
gauge invariance for a Compton-type amplitude to twist-four accuracy.
We have given as well the exact Lorentz structure of the amplitude
motivated by our results at leading order in the coupling constant.
Obviously, beyond leading order, e.g.\ different Lorentz structures
in ${\cal T}^{(1)}_{\mu\nu}$ will be multiplied by independent
functions (see Ref.\ \cite{BelSch98}), analogues of $F_1$ and $F_2$ in
the forward scattering. Our analysis demonstrates a necessity of
introduction of twist-three two-particle generalized distributions.
The latter can be related by means of QCD equation of motion and Lorentz
invariance to the familiar twist-two ones and interaction dependent
antiquark-gluon-quark correlation functions. The former ones represent
a generalization of the Wandzura-Wilczek type relation, while the
last ones give dynamical twist-three contributions. With the assumption
of smallness of dynamical contributions the Wandzura-Wilczek part with
parametrization of leading twist skewed parton distributions can serve
as a model for `transverse' functions. Contributions of these functions
to diverse asymmetries \cite{BelMulNieSch00} will be considered elsewhere
\cite{BelKirMueSch00}.

\vspace{0.5cm}

{\bf Note added}: Recently there appeared a note \cite{PenPolShuStr00}
where the quark hand-bag diagram is calculated in parton model with
transverse momentum being kept. This result overlaps with a part of
our analysis in section \ref{Amplitude}.

\vspace{0.5cm}

We thank A. Kirchner for a collaboration at an early stages of the work.
A.B. would like to thank A. Sch\"afer for the hospitality extended
to him at the Institut f\"ur Theoretische Physik, Universit\"at
Regensburg. This work was supported by Alexander von Humboldt Foundation,
in part by National Science Foundation, under grant PHY9722101 (A.B.)
and by DFG and BMBF (D.M.).


\begin{thebibliography}{99}
\bibitem{MueRobGeyDitHor94}
D. M\"uller, D. Robaschik, B. Geyer, F.-M. Dittes, J. Ho\v{r}ej\v{s}i,
Fortschr. Phys. 42 (1994) 101.
\bibitem{Ji97}
X. Ji, Phys. Rev. D 55 (1997) 7114.
\bibitem{Rad96}
A.V. Radyushkin, Phys. Rev. D 56 (1997) 5524.
\bibitem{DieGouPirTer98}
M. Diehl, T. Gousset, B. Pire, O.V. Teryaev, Phys. Rev. Lett. 81 (1998)
1782.
\bibitem{ColFre98}
J.C. Collins, A. Freund, Phys. Rev. D 59 (1999) 074009.
\bibitem{JiOsb98}
X. Ji, J. Osborne, Phys. Rev. D 58 (1998) 094018.
\bibitem{Fre00}
A. Freund, Phys. Rev. D 61 (2000) 074010.
\bibitem{ItzZub80}
C. Itzykson, J. Zuber, Quantum Field Theory, McGraw-Hill, (New York, 1980).
\bibitem{AniPirTer00}
I.V. Anikin, B. Pire, O.V. Teryaev, {\it On the gauge invariance of the
DVCS amplitude}, hep-ph/0003203.
\bibitem{BelSch98}
A.V. Belitsky, A. Sch\"afer, Nucl. Phys. B 527 (1998) 235.
\bibitem{BelMulNieSch00}
A.V. Belitsky, D. M\"uller, L. Niedermeier, A. Sch\"afer, {\it Leading
twist asymmetries in deeply virtual Compton scattering}, hep-ph/0004059.
\bibitem{BalBra89}
I.I. Balitsky, V.M. Braun, Nucl. Phys. B 311 (1989) 541.
\bibitem{BalBraKoiTan98}
P. Ball, V.M. Braun, Y. Koike, K. Tanaka, Nucl. Phys. B 529 (1998) 323.
\bibitem{GeyLaz99}
B. Geyer, M. Lazar, D. Robaschik, Nucl. Phys. B 559 (1999) 339;\\
B. Geyer, M. Lazar, {\it Twist decomposition of nonlocal light cone
operators. 2. General tensors of 2nd rank}, hep-th/0003080.
\bibitem{BelKirMueSch00}
A.V. Belitsky, A. Kirchner, D. M\"uller,  A. Sch\"afer, {\it
in preperation.}
\bibitem{DieGouPirRal97}
M. Diehl, T. Gousset, B. Pire, J.P. Ralston, Phys. Lett. B411 (1997) 193.
\bibitem{SLAC00}
E143 Collaboration, K. Abe et al., Phys. Rev. Lett. 76 (1996) 587;\\
E155 Collaboration, P.L. Anthony et al., Phys. Lett. B 458 (1999) 529.
\bibitem{Lattice}
M. G\"ockeler, P. Rakow, private communication;\\
M. G\"ockeler, R. Horsley, W. K\"urzinger, H. Oerlich, P. Rakow, G.
Schierholz, {\it The polarized structure function $g_2$: A lattice
study revisited}, hep-ph/9909253, in {\it Polarized Protons at High
Energies --- Accelator Challanges and Physics Opportunities}.
\bibitem{BelMul98}
A.V. Belitsky, D. M\"uller, Nucl. Phys. B 537 (1999) 397;\\
A.V. Belitsky, A. Freund, D. M\"uller, Nucl. Phys. B 574 (2000) 347.
\bibitem{BukFroLipKur85}
A.P. Bukhvostov, G.V. Frolov, L.N. Lipatov, E.A. Kuraev, Nucl. Phys.
B 258 (1985) 601.
\bibitem{Bel00}
A.V. Belitsky, {\it Integrability of twist-three evolution equations in QCD},
hep-ph/0007013.
\bibitem{BraDerMan98}
V.M. Braun, S.E. Derkachov, A.N. Manashov, Phys. Rev. Lett. 81 (1998) 2020.
\bibitem{Bel99}
A.V. Belitsky, Phys. Lett. B 453 (1999) 59; Nucl. Phys. B 558 (1999)
259; Nucl. Phys. B 574 (2000) 407.
\bibitem{DerKorMan99}
S.E. Derkachov, G.P. Korchemsky, A.N. Manashov, Nucl. Phys. B 566 (2000) 203.
\bibitem{AliBraHil91}
A. Ali, V.M. Braun, G. Hiller, Phys. Lett. B 266 (1991) 117.
\bibitem{GeyMueRob96bBelMue97Mue97}
B. Geyer, D. M\"uller, D. Robaschik, {\it The evolution of the nonsinglet
twist-3 parton distribution function}, hep-ph/9611452; \\
A.V. Belitsky, D. M\"uller, Nucl. Phys. B 503 (1997) 279; \\
D. M\"uller, Phys. Lett. B 407 (1997) 314.
\bibitem{PenPolShuStr00}
M. Penttinen, M.V. Polyakov, A.G. Shuvaev, M. Strikman, {\it DVCS
amplitude in the parton model}, hep-ph/0006321.
\end{thebibliography}
\end{document}